\documentclass[reprint,amsmath,amssymb,aps,pre,showpacs]{revtex4-1}
\usepackage{graphicx}
\usepackage{dcolumn}
\usepackage{bm}
\usepackage{hyperref}
\usepackage{bookmark}
\usepackage{subfigure}

\begin{document}
\title{Synchronization of nearly-identical dynamical systems. II Optimized networks}
\author{Suman Acharyya}
\email{suman@prl.res.in}
\author{R. E. Amritkar}
\email{amritkar@prl.res.in}
\affiliation{Physical Research Laboratory, Ahmedabad, India}

\begin{abstract}
In this paper we use the master stability function (MSF) for nearly identical dynamical systems obtained in the previous paper [Phys. Rev. E] to construct optimized networks (ONs) which show better synchronizability. Nearly identical nature is the result of having some node dependent parameters (NDPs) in the dynamics. We study the correlation between various network properties and the values of NDPs on different nodes for the optimized networks and compare them with random networks using the example of coupled R\"ossler systems. In an ON, the nodes with NDP values at one extreme, e.g. nodes with higher frequencies in coupled R\"ossler systems, have higher degrees and are chosen as hubs. These nodes also show higher betweenness centrality. The links in ON are preferably between nodes with large differences in NDP values. The ONs have in general higher clustering coefficient. We also study other network properties such as average shortest path, degree mixing etc. and their relation to the NDP in ON. We 
consider cases of both one and two NDPs and also directed networks.
\end{abstract}


\maketitle

\section{\label{Introduction}Introduction}

The formalism of master stability function (MSF) was proposed by Pecora and Carroll to study the stability of synchronization of coupled identical dynamical systems on different networks on a unified platform. In Part I \cite{previous}, we extend the formalism of MSF to
coupled nearly identical dynamical systems where the different systems have one or more node dependent parameters (NDPs). We apply MSF to study stability of some networks and find that the stability of synchronization can be improved by a suitable realisation of NDP. We demonstrate the improvement of synchronization by considering the example of a star network.
This improvement depends on the structure of the eigenvectors of the coupling matrix. In the ring network, the NDP does not change the synchronization by a significant amount.

The structure of a network plays a major role towards the synchronization of dynamical systems on networks. As discussed in the introduction of Ref.~\cite{previous}, several studies of comparing the synchronization properties of different networks of coupled identical systems have been carried out.
The scale free networks and the small world networks are shown to enhance the synchronizability for coupled identical dynamical systems and this may be due to their smaller shortest path lengths ~\cite{Barahona2002,FernandezPRL2000}. But, other network parameters also play a role in synchronization. It has been found that keeping the path lengths unchanged and increasing the degree heterogeneity reduces the synchronizability of scale free networks~\cite{Nishikawa2003}. This reduction can be overcome by adding suitable weights to the edges of the networks~\cite{MotterEPL2005,MotterPRE2005,ZhouPRL2006,ChavezPRL2005}.

The study of the relation between the network structure and synchronizability for coupled identical systems has motivated us to study a similar problem for coupled nearly identical systems. One such study is reported in \cite{previous} where we show that synchronizability of a network of coupled nearly identical systems can be better than that for coupled identical systems. This is illustrated by using the example of a star network where we show that the critical value of the number of nodes for size instability of synchronization of coupled identical systems can be enhanced by a judicious choice of values of NDP for different nodes. In this paper, we take another approach to the problem, i.e. to construct a network  showing optimized synchronizability. We will refer to this as an optimized network (ON).

For coupled identical systems, Donetti, Hurtado, and M. A. Mu\~{n}oz \cite{Donetti2005} have constructed optimized networks using MSF showing better synchronizability. These ONs are found to have an interwoven structure with a narrow distribution of degree, and of betweenness centrality.
Also, it has been observed that the disassortative mixing of degrees of nodes (negative values of the assortative coefficient) increases the synchronizability of network of coupled identical systems \cite{Bernardo2005,Sorrentino2007}. But, as the disassortative mixing becomes larger, there exists a threshold of the assortative coefficient below which the network loses its synchronizability \cite{ChavezPRE2006}.

When we consider coupled nearly identical systems, a new interesting aspect emerges since the nodes can be identified by their parameter value. Thus the question of relevance is ``what is the relation between the network properties and the NDP in the ON?" E.g. which nodes are likely to be hubs, i.e. have large degree and which nodes are likely to have smaller degree. Many similar questions can be raised in relation to other network parameters and we study them in this paper. Some of the preliminary results of the present study were earlier reported in~\cite{Acharyya2012}.

In this paper, we use the MSF formalism developed in \cite{previous} to construct an ON showing better synchronizability for coupled nearly identical systems and we study the properties of the ON.
In particular, we study the effect of this optimization on various network features such as degree distribution, centrality, shortest path etc.

\section{\label{optnet} Construction of optimized network}

Consider a network of $N$ coupled dynamical systems as
\begin{equation}
\dot{x}_i = f(x_i,r_i) + \varepsilon\sum_{j=1}^N g_{ij} h(x_j);\; i=1,...,N
\label{N-systems}
\end{equation}
where $x_i(\in R^m)$ is the $m$-dimensional state vector of system $i$, $r_i$ is the parameter which makes the systems nonidentical, $f:R^m \rightarrow R^m$ and $h: R^m \rightarrow R^m$ give respectively the dynamical evolution of a single system and the coupling function, $G=[g_{ij}]$ is the coupling matrix and $\varepsilon$ is the coupling constant. The diagonal element of the coupling matrix are $g_{ii}= - \sum_{j\neq i} g_{ij}$. Thus, the coupling matrix satisfies the condition $\sum_{j}g_{ij}=0$ which fulfills the condition for invariance of the synchronization manifold~\cite{Pecora1998}. Let the parameter $r_i = \tilde{r}+\delta r_i$, where $\tilde{r}$ is some typical value of the parameter and $\delta r_i$ is a node dependent small mismatch. We will refer to $r_i$ as the node dependent parameter (NDP).

In \cite{previous}, we study the stability properties of the synchronization of the above system of coupled nearly identical systems. We find that the stability problem can be reduced to two parameters, the network parameter $\alpha$ and the mismatch parameter $\nu_r$. We derive the master stability equations as
\begin{equation}
\dot{\phi} = [D_x f + \alpha D_x h + \nu_r D_r D_x f] \phi
\label{master-stability-equation}
\end{equation}
where $\phi$ is related to the small deviations from synchronization. From the master stability equation we obtain the master stability function (MSF) as the largest Lyapunov exponent of Eq.~(\ref{master-stability-equation}) as a function of $\alpha$ and $\nu_r$.

\begin{figure}
\begin{center}
\includegraphics[width=.9\columnwidth]{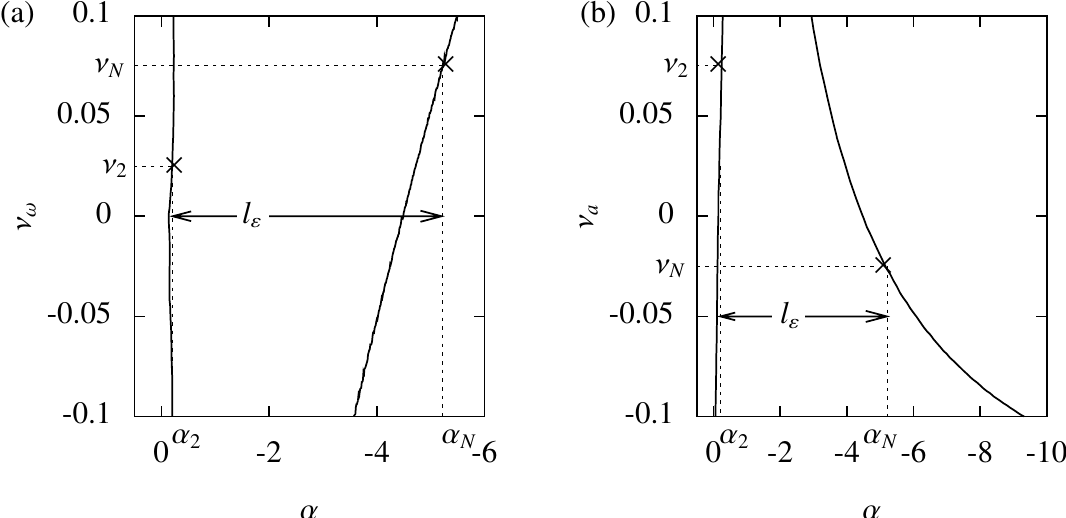}
\end{center}
\caption{\label{msfstablerange} (a) The zero contours of the MSF for coupled nearly identical R\"ossler systems with the frequency $\omega_i$ as the NDP, are plotted on the $\alpha$-$\nu_{\omega}$ plane. Here we take the coupling matrix to be symmetric so that $\alpha$ and $\nu_{\omega}$ are real. The points $(\alpha_2, \nu_2)$ and $(\alpha_N, \nu_N)$ are schematically shown as corresponding to the eigenvalues $\gamma_2$ and $\gamma_N$ of the coupling matrix $G$. The range $l= \alpha_N/\gamma_N - \alpha_2 /\gamma_2$ gives the range of $\varepsilon$ values for stable synchronization. The other R\"ossler parameters are $a=b=0.2,c=7.0$. (b) The zero contours of the MSF for coupled nearly identical R\"ossler systems with parameter $a$ as NDP, are plotted on the $\alpha$-$\nu_a$ plane. Other details are as in (a) with $\omega = 1.0$.}
\end{figure}

As an example we consider coupled R\"ossler systems
\begin{eqnarray}
\dot{x_i} &=& -\omega_i y_i - z_i + \varepsilon \sum_j G_{ij} (x_j - x_i),\nonumber\\
\dot{y_i} &=& \omega_i x_i + a y_i, \label{rossler-system} \\
\dot{z_i} &=& b + z_i(x_i-c), \nonumber
\end{eqnarray}
where $\omega_i$ is the frequency parameter and also NDP. Using Eq.~(\ref{master-stability-equation}) we obtain MSF and Fig.~\ref{msfstablerange}a shows the plot of zero contour of MSF as a function of the two parameters $\alpha$ and $\nu_{\omega}$. In this figure, the stability region of synchronization is bounded by two solids lines and is of V-shape. The width of the stability region increases with the mismatch parameter $\nu_{\omega}$.

An opposite behaviour is observed if in Eqs.~(\ref{rossler-system}), the parameter $a$ becomes the NDP. In this case, the R\"ossler equations are
\begin{eqnarray}
\dot{x_i} &=& -\omega y_i - z_i + \varepsilon \sum_j G_{ij} (x_j - x_i),\nonumber\\
\dot{y_i} &=& \omega x_i + a_i y_i, \label{rossler-system-a} \\
\dot{z_i} &=& b + z_i(x_i-c), \nonumber
\end{eqnarray}
Fig.~\ref{msfstablerange}b shows the plot of zero contour of MSF as a function of the two parameters $\alpha$ and $\nu_a$. The stability region of synchronization is again bounded by two solids lines but is of an inverted V-shape, i.e. $\Lambda$-shape. Thus, the width of the stability region decreases with the mismatch parameter $\nu_a$.

The other two parameters $b$ and $c$ of R\"ossler system do not affect the synchronization region significantly \cite{Acharyya2012}, and hence we will not consider the node dependence of these parameters.

\subsection{\label{opt-method} The optimization method}

 Now, we discuss the problem of constructing optimized networks (ONs) for better synchronizability for a fixed number of links and nodes. We start with a connected random network of $N$ nodes and $E$ links. Let the coupling matrix of this initial random network be $G$. The eigenvalues of the coupling matrix $G$ are $\gamma_1=0>\gamma_2\geq\gamma_3\geq\cdots\gamma_N$. Our procedure consists of rewiring the links to get the optimal network. For rewiring we use the Metropolis algorithm. For each $\gamma_i$, we can find the value of the mismatch parameter $\nu_i$ using the eigenvectors of the coupling matrix $G$. The MSF allows us to determine the corresponding $\alpha_i$ values. Thus, we can determine $l=\alpha_N/\gamma_N \sim \alpha_2/\gamma_2$ as the interval of the coupling constant $\varepsilon$ for stable synchronization. This is shown schematically in Fig.~\ref{msfstablerange}.

In the initial network we randomly delete an existing link and create a new link at a link vacancy making sure that the new network is connected. Let, the coupling matrix of the resultant network be $G'$. We determine the interval of stable synchronization, say $l'=\alpha^{'}_N/\gamma^{'}_N \sim \alpha^{'}_2/\gamma^{'}_2$, for $G'$ using the procedure described above. When $l' > l$ we accept the new network $G'$, else we accept it with a probability $e^{(l' - l)/T}$, where $T$ is a temperature like parameter. This rewiring procedure defines one Monte Carlo step of our procedure and is repeated several times. We use a slow annealing procedure. $T$ is kept fixed for 1000N Monte Carlo steps or 10N accepted ones, whichever occurs first. Then $T$ is decreased by $10\%$ so that stimulated annealing or slow cooling occurs~\cite{Binder_book,Binder_book_2}. When the stability interval becomes approximately constant over successive annealing steps, we assume that a reasonably good approximation for the optimal 
topology has been found.

\begin{figure}
\begin{center}
\includegraphics[width=.9\columnwidth]{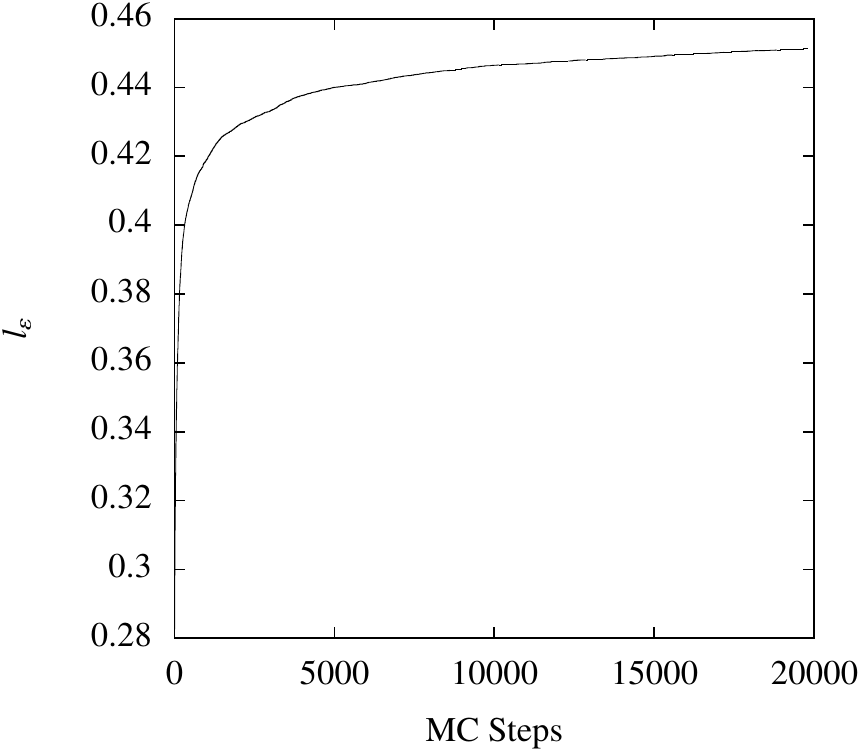}
\end{center}
\caption{\label{l} The stable interval $l_{\varepsilon}$ is plotted as a function of Monte Carlo steps for a directed network of $32$ coupled R\"ossler systems which have $\omega$ as NDP.}
\end{figure}

In Fig.~\ref{l} the typical increase of the stable interval $l_{\varepsilon}$ is shown as a function of the Monte Carlo steps for a directed network of $32$ coupled R\"ossler systems with NDP $\omega$. The stable interval rises sharply and then saturates to a higher value when optimal topology is found.

\begin{figure*}
\subfigure[Initial random network]{
\centering
\includegraphics[scale=.25]{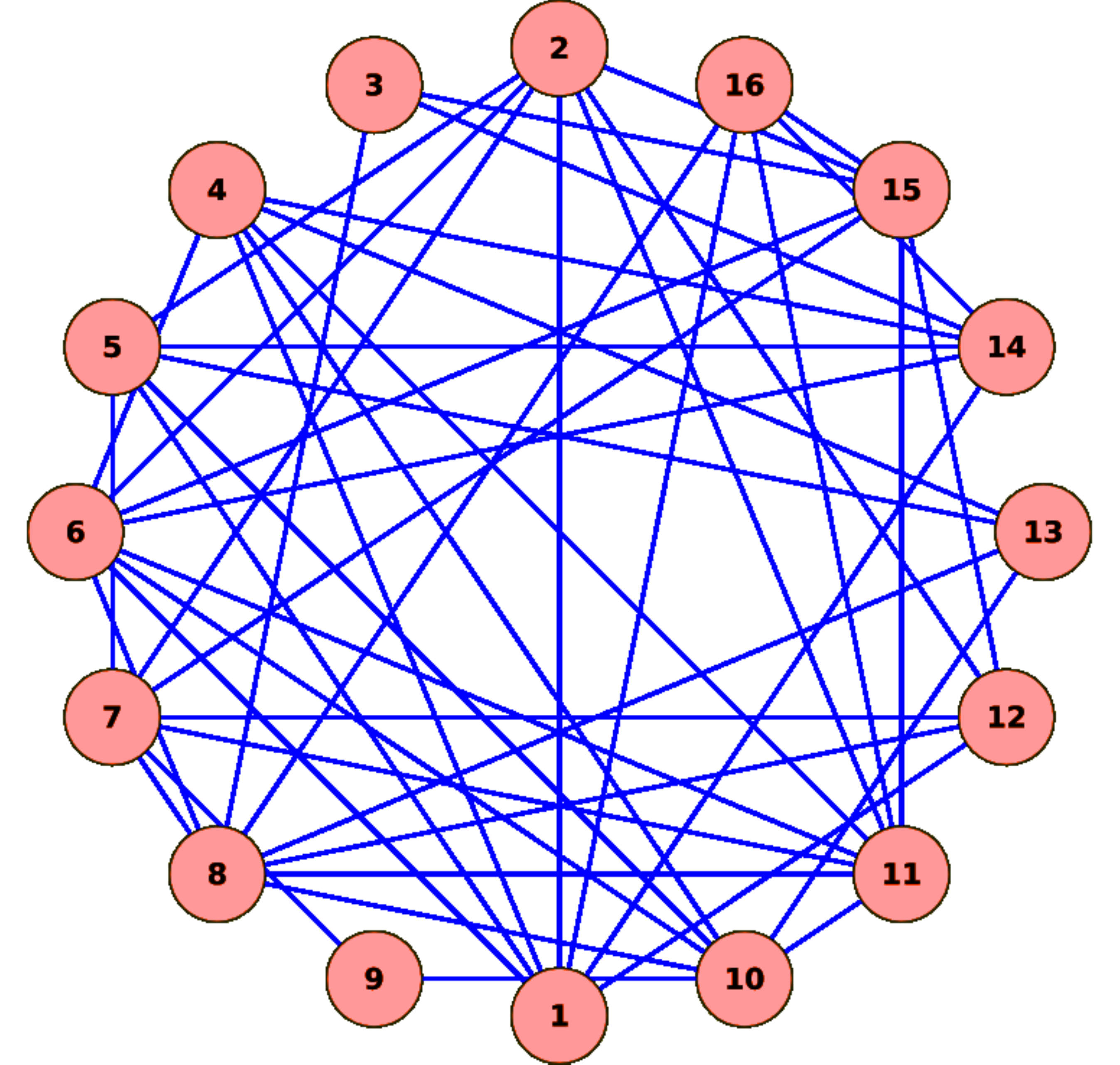}
}
\hspace{20pt}
\subfigure[Optimized network]{
\centering
\includegraphics[scale=.25]{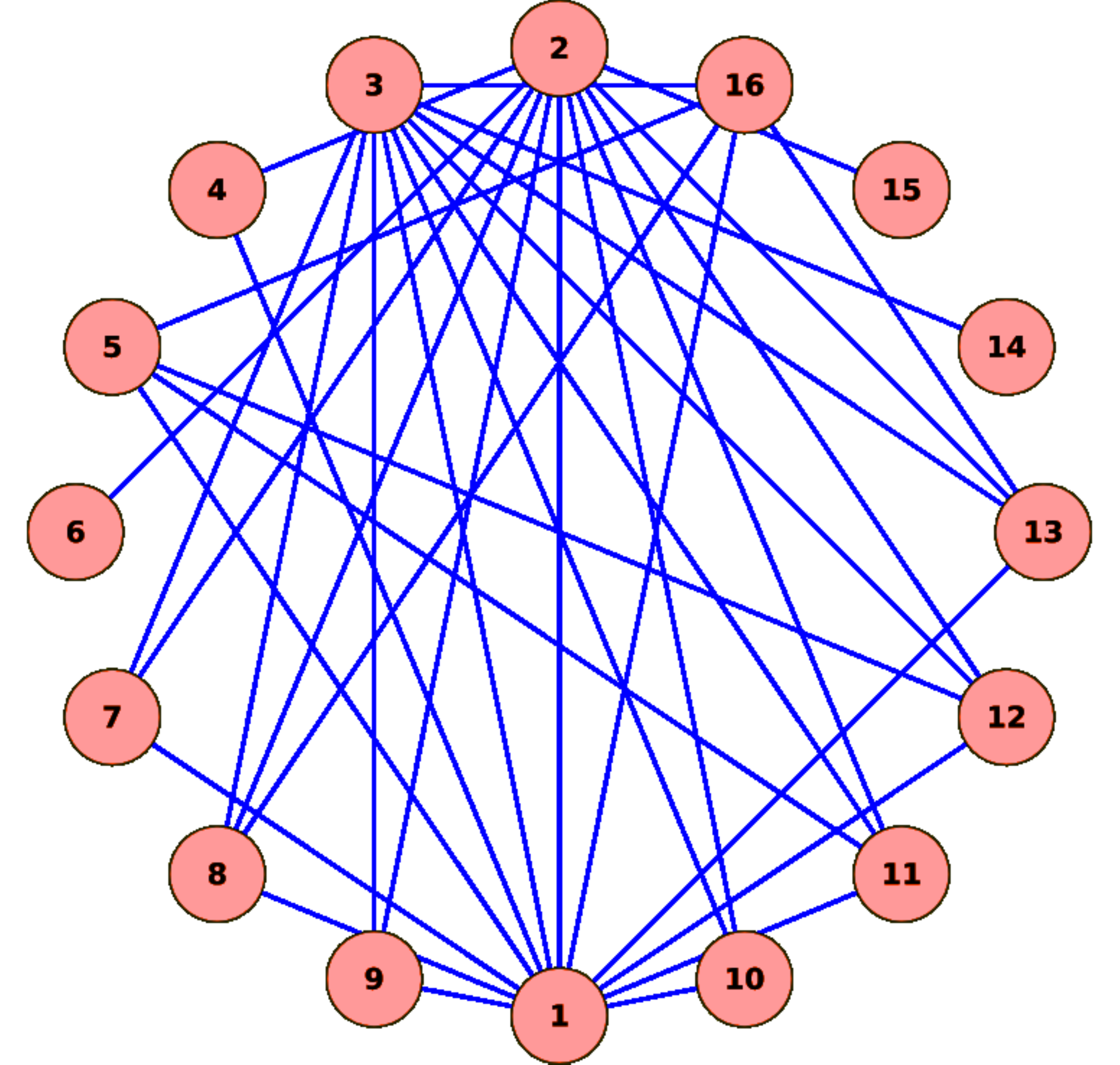}
}
\caption{\label{ini_opt_net}(a) Shows a random undirected network of $16$ coupled R\"ossler systems with NDP $\omega$. The NDP $\omega$ is distributed uniformly in the interval $(1.01,0.99)$, and node labeled $(1)$ has the highest parameter value and the other nodes are labeled in a decreasing order of $\omega$. (b) Shows the synchronization optimized network constructed from the random network of (a).}
\end{figure*}

Fig.~\ref{ini_opt_net}(a) shows a random undirected network with $16$ coupled R\"ossler systems and a total $48$ edges. The coupled R\"ossler systems has NDP $\omega$ which is distributed uniformly in the interval $(1.01,0.99)$. The  nodes $(1)$ to $(16)$ are numbered in the descending order of NDP $\omega$. Node $(1)$ has the maximum value $\omega=1.01$ and node $(16)$ has the minimum value $\omega=0.99$. In Fig.~\ref{ini_opt_net}(b) a synchronization optimized network is shown which is constructed from the random network shown in Fig.~\ref{ini_opt_net}(a). We note that in the optimized network the nodes with higher value of $\omega$ have more connections than the other nodes, a property which is discussed in Section~\ref{Degree of Nodes}.

\section{\label{properties} Properties of optimized network}

We now study the properties of optimized network, and in particular, try to find the relation between the NDP and the structural properties. For this study we use the system of coupled R\"ossler oscillators given by Eq.~(\ref{rossler-system}) where the frequency $\omega_i$ is the NDP and Eq.~(\ref{rossler-system-a}) where the parameter $a$ is the NDP.

For numerical simulations we consider $N$ coupled R\"ossler systems where $N=32, 64$ or $256$. We start with a random network where the edges are randomly connected with some probability. We evolve the network using the optimization procedure described in the previous section and obtain an optimized network. The network properties are obtained by averaging over 100 realizations of the entire procedure. The R\"ossler parameter used in these calculations are $b = 0.2,\, c=7.0$, and $a=0.2$ when the frequency $\omega$ is NDP and $\omega=1.0$ when $a$ is NDP.

In this section we consider undirected networks. The case of directed networks is treated separately.

\subsection{\label{Degree of Nodes}Degree of nodes}

\begin{figure}
\begin{center}
\includegraphics[width=.9\columnwidth]{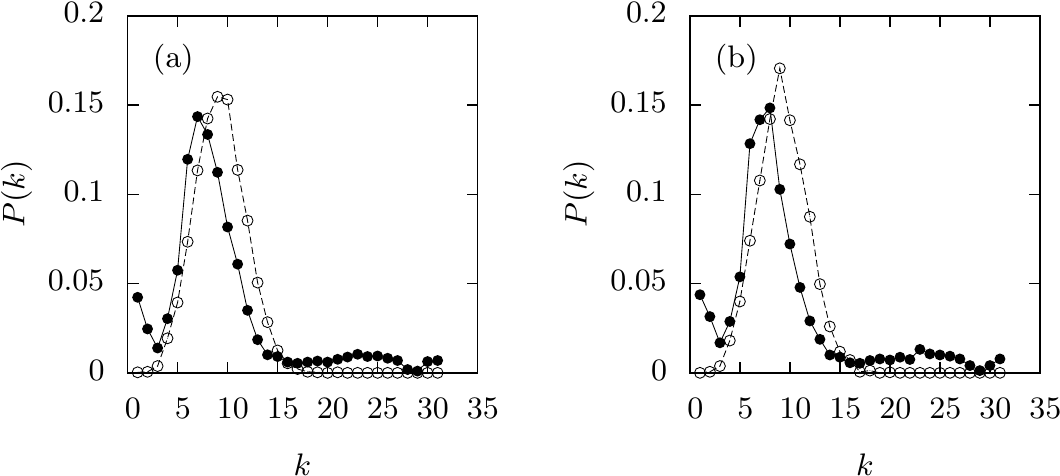}
\end{center}
\caption{\label{p_dist_omega} (a) The degree distributions $P(k)$ as a function of $k$ for the initial random network (open circles) and the optimized network (solid circles) are shown for $32$ coupled R\"ossler systems with $\omega$ as NDP. We can see that the degree distribution of the random network is Gaussian and has a single peak. The major part of the gaussian is retained in the degree distribution $P(k)$ of the optimized network, but we have two additional features, a small broad peak for larger degrees showing formation of a few hubs and another peak for very small degrees. (b) The degree distributions $P(k)$ as a function of $k$ for the initial random network (open circles) and the optimized network (solid circles) are shown for $32$ coupled R\"ossler systems as in (a), but with $a$ as NDP.}
\end{figure}

One of the basic topological characterization of a network is the degree distribution $P(k)$, the probability that a randomly chosen node has degree $k$. In Fig.~\ref{p_dist_omega}a we compare the degree distribution of the initial random network (open circles) and the optimized network (solid circles) with the frequency $\omega$ as NDP. The initial random network has the expected gaussian distribution. In the optimized network, the major part of the degree distribution retains almost similar shape to that of the random network, but we have a small hump for large degrees corresponding to the creation of a few hubs. In addition we have a small increase in the probability for smaller degree nodes. This behavior is surprisingly different from that observed for identical systems where for optimized networks the standard deviation of the degree distribution decreases with optimization \cite{Donetti2005}.

In Fig.~\ref{p_dist_omega}b we show the degree distribution of the initial random network (open circles) and the optimized network (solid circles) with parameter $a$ as NDP. We see from Fig.~\ref{p_dist_omega}b that the behavior of degree distribution is very similar to that with the frequency $\omega$ as NDP.

To understand the degree distribution, we investigate the relation between the NDP values and the corresponding degrees.
 We ask the question, ``Do some nodes preferentially become the hubs depending on NDP and some other nodes have smaller degrees?". To this end, we define the correlation between the degree of a node and its NDP as
\begin{equation}
\rho_{rk} = \frac{\langle(r_i - \langle r_i\rangle)(k_i - \langle k_i\rangle)\rangle}{\sqrt{\langle(r_i - \langle r_i\rangle)^2\rangle\langle(k_i - \langle k_i\rangle)^2\rangle}}
\label{param-deg-corr}
\end{equation}
where, $k_i=-g_{ii}$ is the degree of node $i$ and $r_i$ is its NDP. For a random network $\rho_{rk}=0$.

\begin{figure}
\centering
\includegraphics[width=.9\columnwidth]{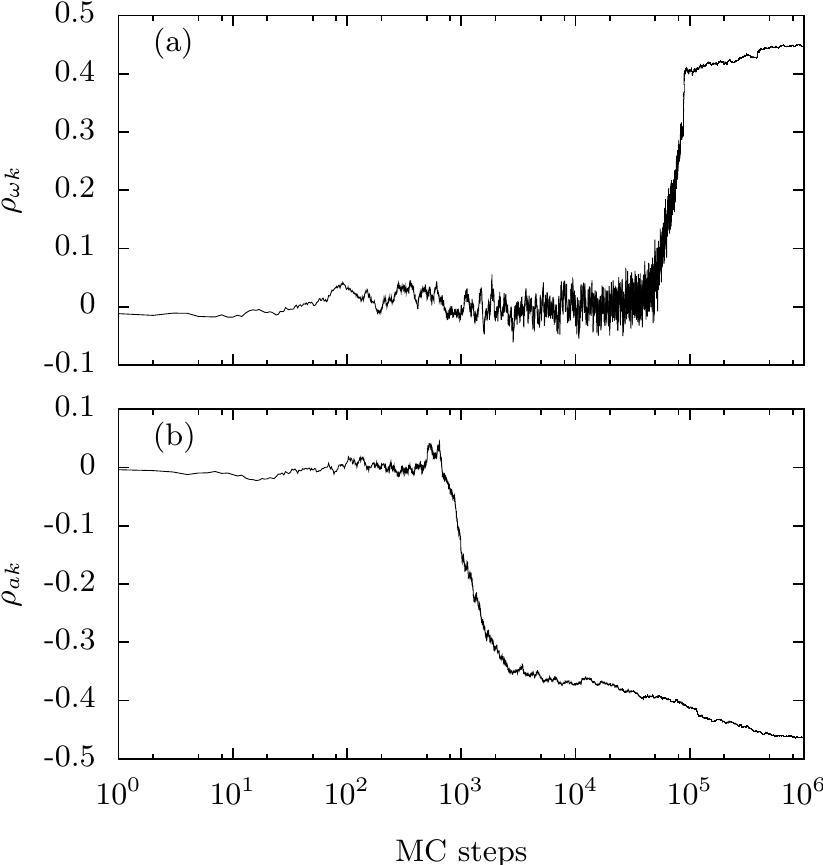}
\caption{\label{rho_omegak} (a) The correlation coefficient $\rho_{\omega k}$ between the NDP $\omega$ and the degree for coupled R\"ossler systems is plotted as a function of Monte Carlo steps. The correlation coefficient increases from zero and saturates to a positive value. (b) The correlation coefficient $\rho_{a k}$ between the NDP $a$ and the degree for coupled R\"ossler systems is plotted as a function of Monte Carlo steps. The correlation coefficient decreases from zero and saturates to a negative value.}
\end{figure}

In Fig.~\ref{rho_omegak}a we plot $\rho_{\omega k}$ as a function of the Monte Carlo steps for a system of coupled R\"ossler oscillators with $\omega_i$ as NDP. From this figure we can see that the correlation coefficient $\rho_{\omega k}$, increases from zero and saturates to a positive value. Thus, in the optimized network, the nodes with larger frequency values have higher degree and nodes with smaller frequency have smaller degree. This explains the peaks for large and small degrees observed in the degree distribution (Fig.~\ref{p_dist_omega}a). If we look at the eigenvectors of the coupling matrix $G$, then in the optimized network, the eigenvectors corresponding to the extreme transverse eigenvalues develop a few large components which correspond to nodes with larger and smaller frequencies. These large components lead to an increase in the corresponding mismatch parameters $\nu_2$ and $\nu_N$ and hence to an improved stability \cite{Acharyya2012} (see Fig.~\ref{msfstablerange}).

In Fig.~\ref{rho_omegak}b we plot $\rho_{a k}$ as a function of the Monte Carlo steps for a system of coupled R\"ossler oscillators with $a$ as NDP. From this figure we can see that in the optimized network, the correlation coefficient $\rho_{a k}$, decreases from zero and saturates to a negative value. Here, the nodes with smaller values of $a$ have larger degrees and those with larger values of $a$ have smaller degrees. This behavior is obtained since the the stability region of MSF for $\omega$ as NDP has V-shape while that for $a$ as NDP has $\Lambda$-shape.

\begin{figure}
\begin{center}
\includegraphics[width=.9\columnwidth]{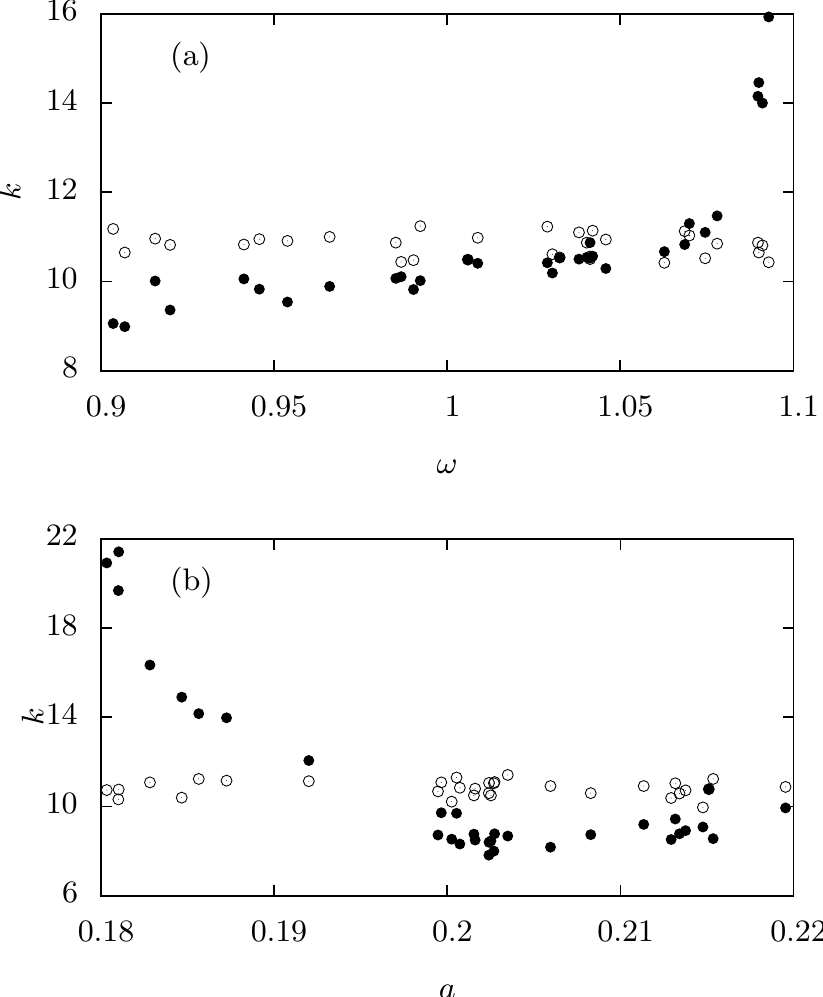}
\end{center}
\caption{\label{OmegaVsDegree} (a) The degree $k$ of the nodes of the random network (open circles) and optimized network (closed circles) are plotted as a function of the NDP $\omega$ for $32$ coupled R\"ossler systems. The nodes with higher $\omega$ value have higher degree. For other frequencies, there is a general decline in the degree of nodes, with somewhat larger decrease for smaller $\omega$ values. (b) The degree $k$ of the nodes of the random network (open circles) and optimized network (closed circles) are plotted as a function of the NDP $a$ for $32$ coupled R\"ossler systems. The nodes with smaller $a$ value have higher degree. For other frequencies, there is a general decline in the degree of nodes, with somewhat larger decrease for larger $a$ values.}
\end{figure}

The relation between the frequency and degree can be made more explicit if we plot the degree of the nodes in the optimized network as a function of the frequency parameter $\omega$ and this is shown in Fig.~\ref{OmegaVsDegree}a. The figure shows that the nodes with higher $\omega$ values have higher degrees. For other frequencies there is a general decrease in the degree with a somewhat larger decrease for smaller frequencies. This is consistent with the behavior of the correlation coefficient $\rho_{\omega k}$ in Fig.~\ref{rho_omegak}a.

In Fig.~\ref{OmegaVsDegree}b, we plot the degree of the nodes in the optimized network as a function of the parameter $a$. The figure shows that the nodes with lower $a$ values have higher degrees. For other frequencies there is a general decrease in the degree with a somewhat larger decrease for larger $a$. This is consistent with the behavior of the correlation coefficient $\rho_{a k}$ in Fig.~\ref{rho_omegak}b.

In this subsection we have studied the behavior of the degree distribution, the NDP-degree correlation coefficient and a plot of degree vs NDP. In all the three cases the behavior for $\omega$ as NDP and $a$ as NDP is statistically similar if we make an equivalence between larger frequencies and smaller $a$ values and between smaller frequencies and larger $a$ values. The origin of this is in the shape of the stability region of MSF which has V-shape for $\omega$ as NDP and $\Lambda$-shape for $a$ as NDP. We have observed that this equivalence holds for other network properties which are studied in the following subsections. Hence, in these subsections we report only the results for $\omega$ as NDP. The behavior with $a$ as NDP can be easily deduced with the above equivalence.

\subsection{Links between nodes}

In the random network the links are assigned randomly with some probability. We now investigate to find whether there is any preference for links depending on the NDPs of the connecting nodes.

We define the correlation between the NDP difference between a pair of nodes and the link between them as
\begin{equation}
\rho_{rA} = \frac{\langle(\delta r_{ij} - \langle\delta r_{ij}\rangle)(A_{ij} - \langle A_{ij}\rangle)\rangle}{\sqrt{\langle(\delta r_{ij} - \langle\delta r_{ij}\rangle)^2\rangle\langle(A_{ij} - \langle A_{ij}\rangle)^2\rangle}}
\label{paramdiff-adj-corr}
\end{equation}
where, $\delta r_{ij} = |r_i - r_j|$ is the NDP difference between node $i$ and node $j$ and $A$ is the adjacency matrix, such that if node $i$ couples with node $j$ then $A_{ij}=1$, otherwise $A_{ij}=0$. We do not consider self loops so the diagonal elements of adjacency matrix are zero $A_{ii}=0$. For a random network $\rho_{rA}=0$.

\begin{figure}
\centering
\includegraphics[width=.9\columnwidth]{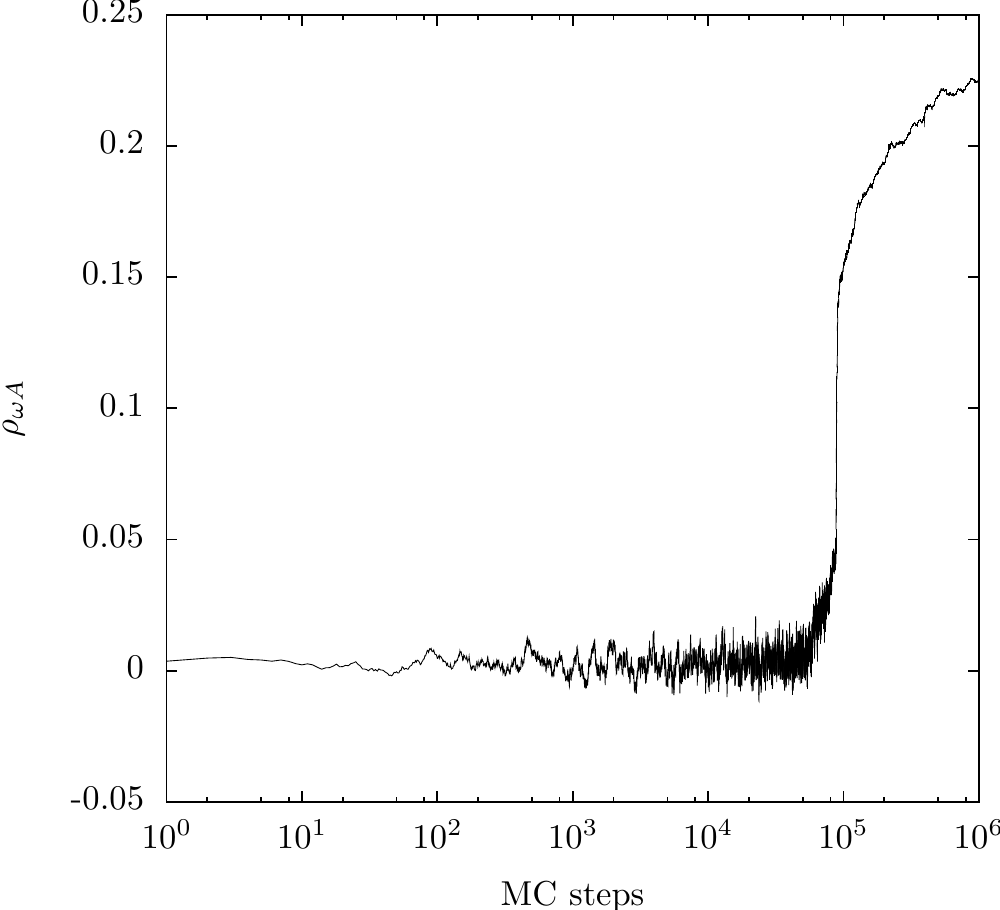}
\caption{\label{rho_omegaA} The correlation coefficient $\rho_{\omega A}$ between the frequency difference of a pair of nodes and their link is plotted as a function of Monte Carlo steps for 32 coupled R\"ossler oscillators. This correction coefficient increases from zero and saturates to a positive value.}
\end{figure}

In Fig.~\ref{rho_omegaA} we plot the correlation coefficient $\rho_{\omega A}$ between the frequency differences of a pair of nodes and their link as a function of Monte Carlo steps for 32 coupled R\"ossler oscillators. From the figure we can see that the correlation coefficient increases from zero and saturates to a positive value. Thus, the pair of nodes with larger parameter difference are preferred for links in the optimized networks. This observation may be explained by noting that connecting nodes at the two extremes of frequencies can help in stabilizing the network in a better way and hence improve synchronization.

\subsection{Clustering coefficients}

The clustering coefficient is a measure of how closely knit a network is and is a local property. It quantifies the possibility that two neighbors of a common node are also neighbors. The clustering coefficient $c_i$ of node $i$ is defined as
\begin{equation}
c_i = \frac{2 e_i}{k_i(k_i-1)}
\end{equation}
where, $e_i$ is the number of edges that exist among the neighbors of node $i$ and $k_i$ is the degree of node $i$.  The clustering coefficient $C$ of the entire network is defined as
\begin{equation}
C = \frac{1}{N}\sum_i c_i.
\end{equation}

\begin{figure}
\centering\includegraphics[width=.9\columnwidth]{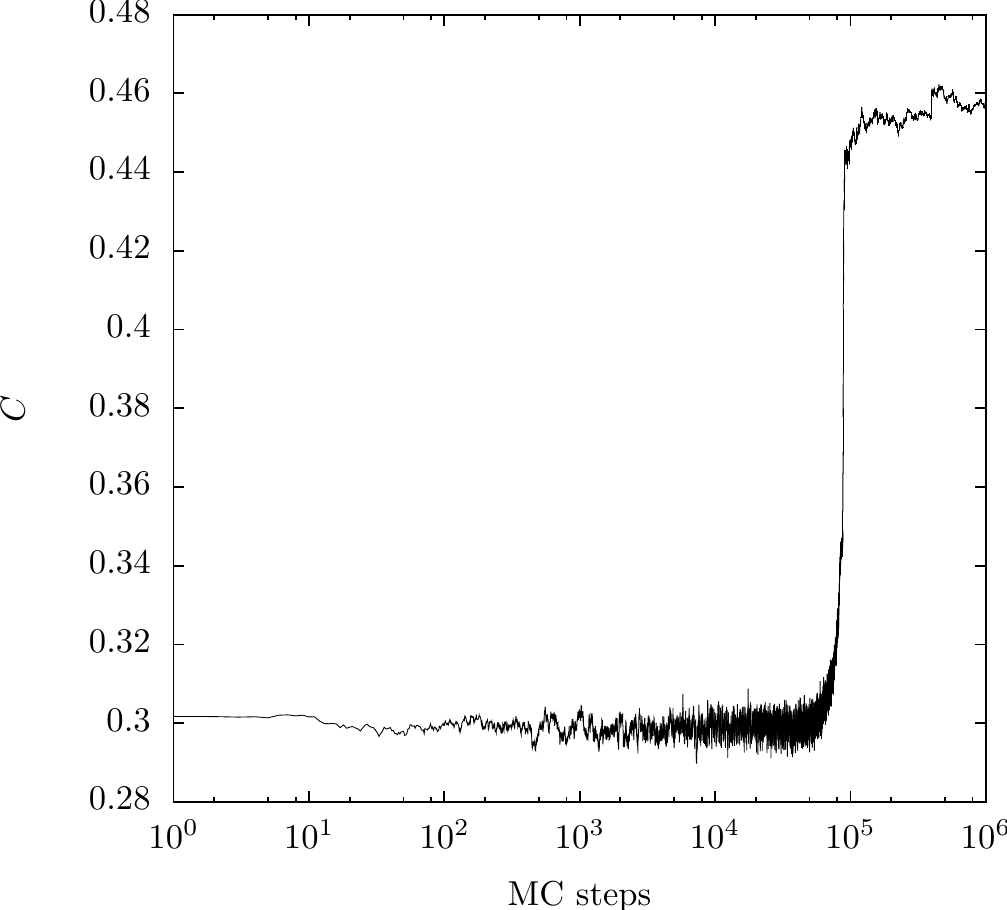}
\caption{\label{AvgAvgClustCoeff_omega} The clustering coefficient $C$ of a network of $32$ coupled R\"ossler systems is plotted as a function of the Monte Carlo steps. The clustering coefficient increases and saturates to a higher positive value.}
\end{figure}

In Fig.~\ref{AvgAvgClustCoeff_omega} the clustering coefficient $C$ of the network is plotted as a function of the Monte Carlo steps. From Fig.~\ref{AvgAvgClustCoeff_omega} we can see that the clustering coefficient of the network increases and saturates to a higher positive value. Thus, the optimized network has more local structure than the random network, i.e. there are more triangles than the random network. The result is intuitively easy to understand. Forming a loop will enhance the stability of synchronization due to a faster feedback and smaller the size of the loop better will be the result. We note that the behavior is similar to that for coupled identical oscillators where it has been noticed that networks with larger value of clustering coefficient have better stability of synchronization \cite{Donetti2005}.

\begin{figure}
\begin{center}
\includegraphics[width=.9\columnwidth]{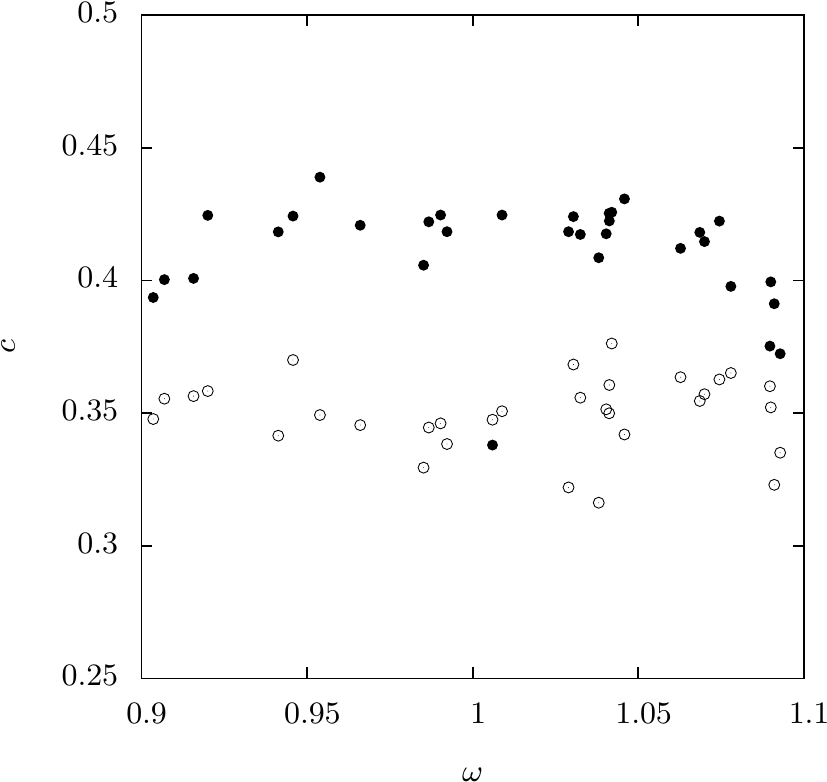}
\end{center}
\caption{\label{ClusteringCoefficient_omega} The clustering coefficients $c$ of individual nodes of the optimized networks are plotted as a function of the NDP $\omega$ for a random network (open circles) and the optimized network (solid circles) for 32 coupled R\"ossler oscillators.}
\end{figure}

In Fig.~\ref{ClusteringCoefficient_omega} the clustering coefficient $c$ of individual nodes of the random (open circles) and optimized networks (closed circles) are plotted as a function of the NDP $\omega$. There is a general increase in the clustering coefficient for the optimized network with a somewhat less increase for larger and smaller frequencies.

\subsection{Average shortest path length}

\begin{figure}
\begin{center}
\includegraphics[width=.9\columnwidth]{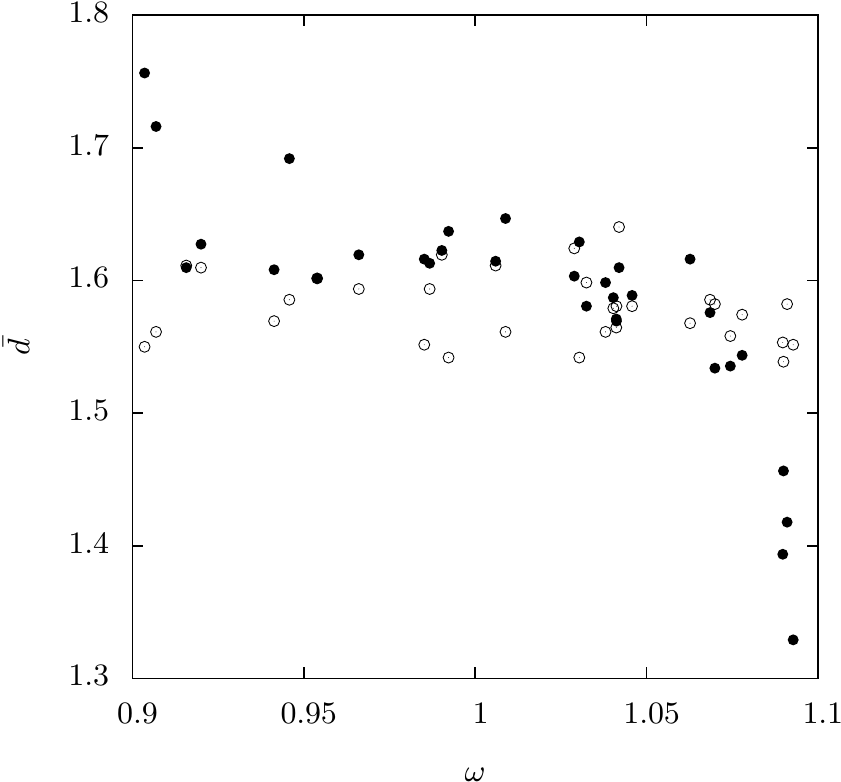}
\end{center}
\caption{\label{AverageShortestPath_omega} The average shortest path lengths $\bar{d}$ of the nodes of the random (open circles) and optimized network (solid circles) are plotted as a function of R\"ossler parameter $\omega$. The nodes with higher parameter values have average shortest path lengths smaller than the average for the entire network.}
\end{figure}

Shortest paths play an important role in the transport and communication within a network. We define the average shortest path length for node $i$ as,
\begin{equation}
\bar{d}_i = \frac{\sum_{j;j\neq i}^N d_{ij}}{N-1}
\label{shortpathlen}
\end{equation}
where, $d_{ij}$ is the shortest path length connecting nodes $i$ and $j$. In Fig.~\ref{AverageShortestPath_omega} the average shortest path length of the individual nodes for the random (open circles) and optimized networks (closed circles) are plotted as a function of NDP $\omega$ for 32 coupled R\"ossler oscillators. From the figure we can see that the shortest path lengths of the nodes with higher $\omega$ values is smaller than the average value. This is the expected behavior since the nodes with higher $\omega$ values have larger degrees (see Sec.~IIIA).

\subsection{Betweenness centrality and closeness centrality}

In this section we discuss two major quantities  which are important for information transformation in a network, the betweenness centrality ($C_B$) and the closeness centrality ($C_C$).

\begin{figure}
\begin{center}
\includegraphics[width=.9\columnwidth]{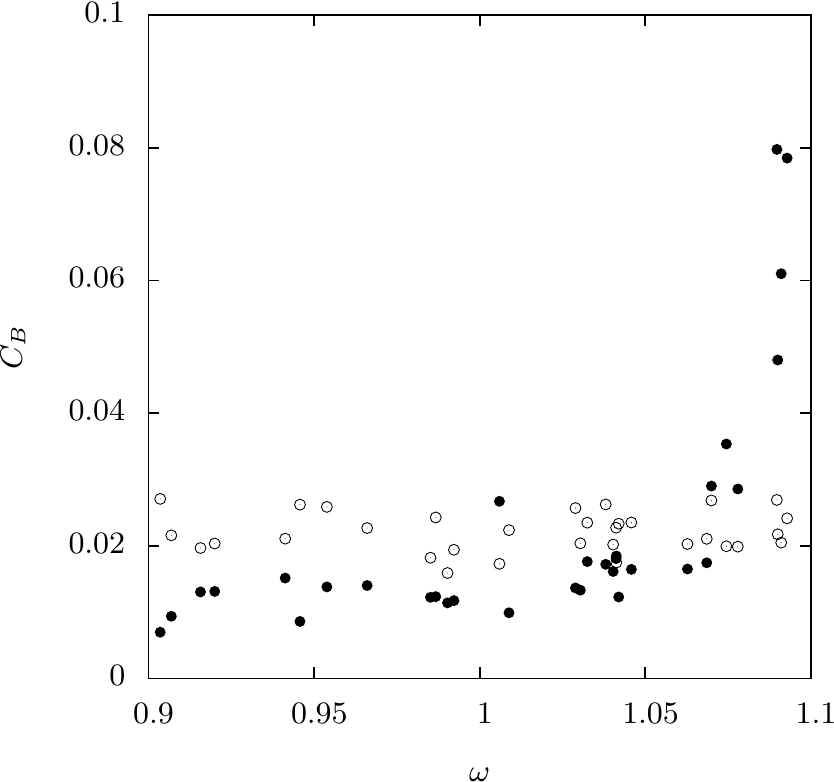}
\end{center}
\caption{\label{BetweennessCentrality_omega} The betweenness centrality $C_B$ of the nodes of the random (open circles) and optimized networks (solid circles)are plotted as a function of the NDP, frequency $\omega$.}
\end{figure}

The betweenness centrality is a measure of the extent to which a node lies on the shortest paths between other nodes and is defined as
\begin{equation}
C_B(i) = \sum_{j,k,j > k}\frac{n_{jk}(i)}{n_{jk}} \label{betweenness-centrality}
\end{equation}
where $n_{jk}$ is the number of shortest paths connecting nodes $j$ and $k$ and $n_{jk}(i)$ is the number of shortest paths connecting $j$ and $k$ and passing through $i$. In Fig.~\ref{BetweennessCentrality_omega} the betweenness centrality $C_B$ of the nodes of the random (open circles) and optimized networks (solid circle) are plotted as a function of the NDP $\omega$, for 32 coupled R\"ossler oscillators. From this figure we can see that the nodes with higher frequency value have larger betweenness centrality and this is consistent with the observation in Sec. IIIA that nodes with larger frequency are more likely to be hubs in the optimized network.

\begin{figure}
\begin{center}
\includegraphics[width=.9\columnwidth]{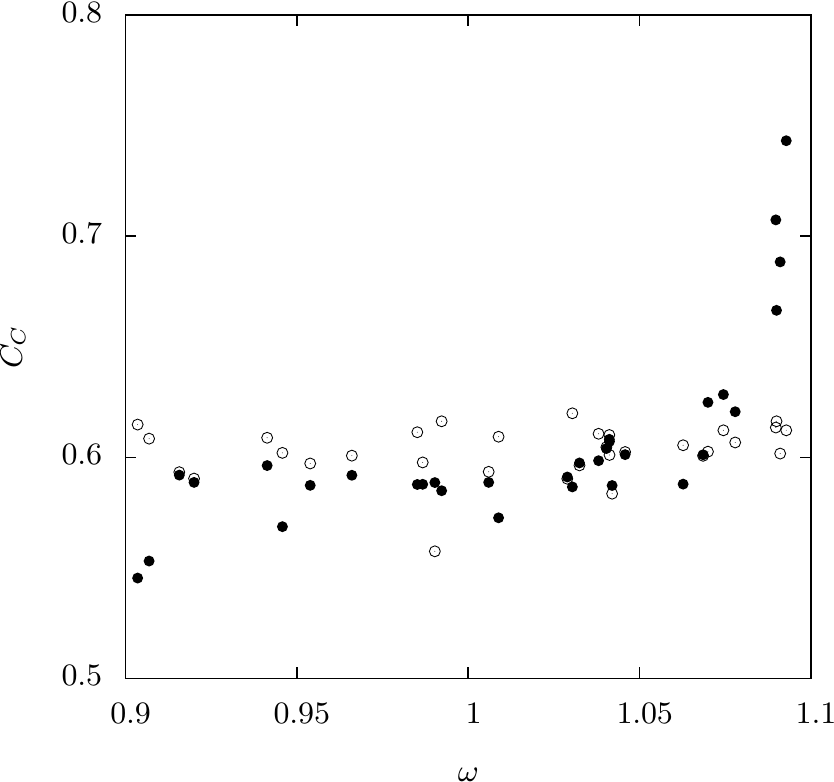}
\end{center}
\caption{\label{ClosenessCentrality_omega} The closeness centrality $C_C$ of the nodes of the random (open circles) and optimized networks (solid circles) are plotted as a function of the R\"ossler parameter $\omega$.}
\end{figure}

The closeness centrality $C_C$ can be regarded as a measure of the efficiency of information transfer on a network and is the inverse of the time taken to spread information from a given vertex to others in the network. The closeness centrality is defined as,
\begin{equation}
C_C(i) = \frac{1}{\sum_j d_{ij}} \label{closeness-centrality}
\end{equation}
where $d_{ij}$ is the shortest distance between nodes $i$ and $j$. In Fig.~\ref{ClosenessCentrality_omega} the closeness centrality $C_C$ of the nodes of the random (open circles) and optimized network (solid circle) are plotted as a function of the R\"ossler parameter $\omega$. From this figure we can see that the nodes with higher frequency value have higher closeness centrality. This is expected since closeness centrality is the inverse of the average shortest path length and average shortest path length is smaller for higher frequency nodes as discussed in Sec. IIID

\subsection{Degree mixing in networks}

In this section we study assortative mixing in networks~\cite{NewmanPRL2002,NewmanPRE2003}. Assortative mixing gives the tendency of nodes to be connected with nodes of comparable degrees. Let the degrees of nodes at the ends of the $i$th edge connecting nodes $j$ and $l$ be $(k_j)_i$ and $(k_l)_i$. Following Ref~\cite{NewmanPRL2002} the assortative coefficient $r$ can be calculated as
\begin{equation}
r = \frac{\frac{1}{M} \sum_i (k_j)_i (k_l)_i - \left[\frac{1}{2M} \sum_i ((k_j)_i+(k_l)_i)\right]^2}{\frac{1}{2M}\sum_i ((k_j)_i^2 + (k_l)_i^2) - \left[\frac{1}{2M} \sum_i ((k_j)_i+(k_l)_i)\right]^2}, \label{assortative_coefficient}
\end{equation}
where, $M$ is the total number of edges in the network and the sums are over all the edges. When comparable degree nodes get connected the correlation coefficient $r$ is positive and the network is called assortative network. The network is called disassortative network when the coefficient $r$ is negative. This happen when high degree nodes are connected with low degree nodes. For networks which show no assortative mixing the correlation coefficient $r$ is zero. The random networks of Erd\H{o}s and R\'{e}nyi and the scale free network model of Barab\'{a}si and Albert shows no assortative mixing. It has been observed that many naturally evolving networks, such as internet, WWW, protein interaction, neural networks, etc. shows disarrortative mixing of degree~\cite{NewmanPRL2002}.

\begin{figure}
\begin{center}
\includegraphics[width=.9\columnwidth]{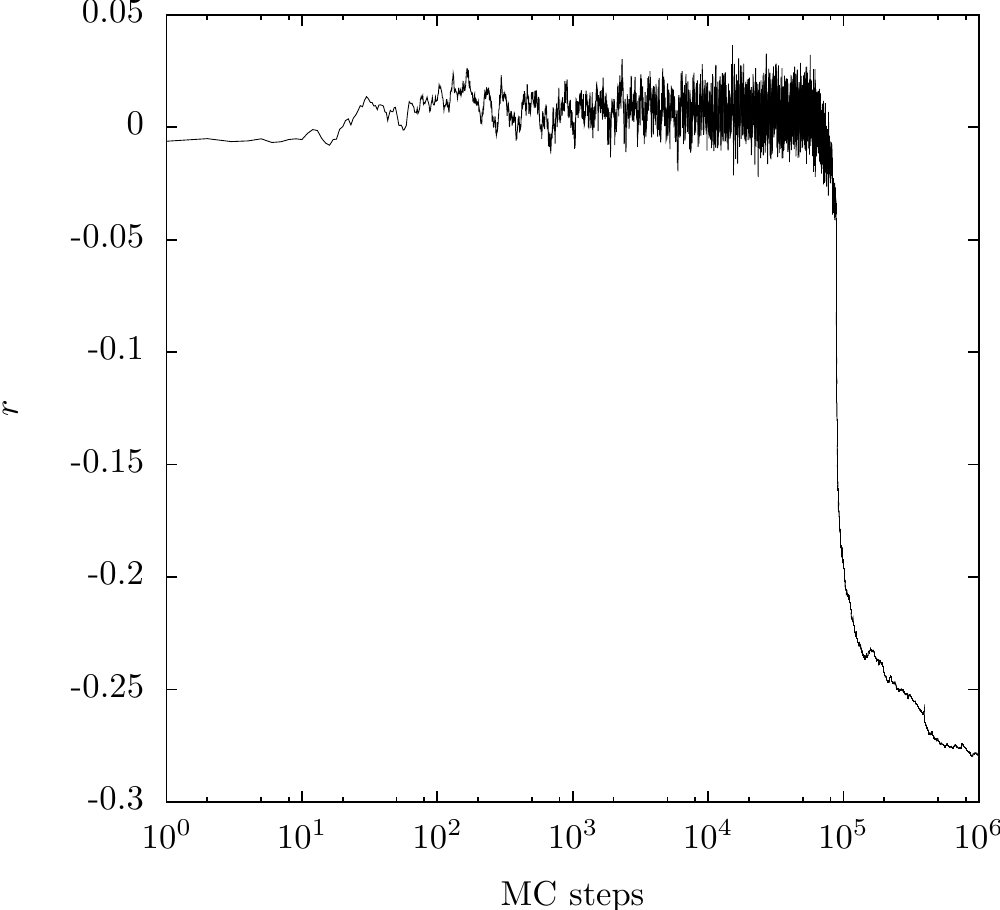}
\end{center}
\caption{\label{AssortativeMixing} The degree mixing coefficient is plotted as a function of Monte Carlo iterations. For some initial steps the coefficient remains near zero, then it decreases and becomes negative.}
\end{figure}

In Fig.~\ref{AssortativeMixing} the assortative correlation coefficient $r$ is shown as a function of Monte Carlo iterations. As we start with an initial random network, the assortative coefficient $r$ remains zero for a few of the initial iterations after which it starts decreasing and becomes negative. Thus the optimized network is disassortative. This behavior is consistent with that of coupled identical systems.

\section{\label{omega-a-mismatch} Optimized network with two parameters as NDPs}

In this section we consider a more general situation where the coupled systems have node dependence in more than one parameter. The master stability equation is \cite{previous}
\begin{equation}
\dot{\phi} = [D_x f + \alpha D_x h + \sum_{k=1}^q \nu_{r_k} D_{r_k} D_x f] \phi
\label{master-stability-equation-q}
\end{equation}
where we have $q$ NDPs $\nu_{r_k}, k=1,\ldots,q$.

Let us consider the example of coupled R\"ossler systems with node dependence in two parameters, $\omega$ and $a$. For this case the MSF is calculated from Eq.~(\ref{master-stability-equation-q}) as a function of $(\alpha,\nu_{\omega},\nu_a)$. The zero contour surfaces of the MSF are shown in Fig.~4 in the three dimensional space, $(\alpha,\nu_{\omega},\nu_a)$. The MSF is negative in the region bounded by the two surfaces in the figure and thus this region gives the stable region of synchronization. The stable region increases with increase in $\nu_{\omega}$ and at the same time it decreases with increase in $\nu_a$.

Now we consider the problem of constructing synchronized optimized network with two NDPs from an initial random network with fixed number of links and nodes. To construct synchronized optimized networks we follow the Monte Carlo optimization method which is discussed in section~\ref{opt-method}.

\subsection{Degree of nodes}

\begin{figure}
\begin{center}
\includegraphics[width=.9\columnwidth]{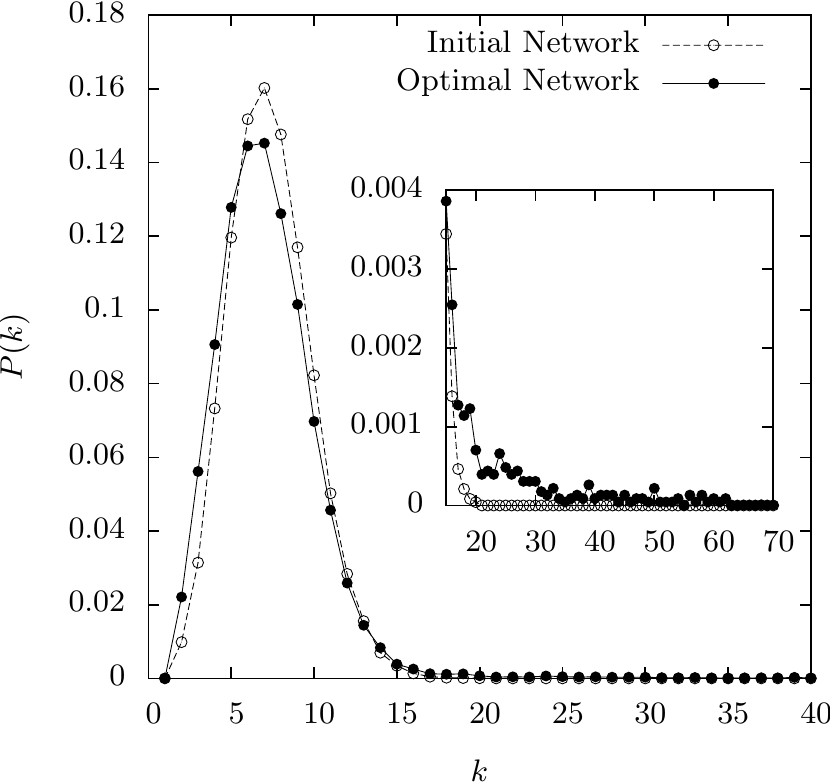}
\end{center}
\caption{\label{DegreeDist_omega_a} The degree distribution $P(k)$ of the initial random (dashed line) and optimized networks (solid line) is shown for $64$ coupled R\"ossler systems. Inset shows the peak at the higher degrees for the optimized network.}
\end{figure}

The degree distributions $P(k)$ of the initial random (dashed line) and optimized network (solid line) with two NDPs are shown in Fig.~\ref{DegreeDist_omega_a}. The degree distribution $P(k)$ of the initial random network has one peak at the average value of degree. While the optimized network has two peaks, one large peak at the average value of degree of the network and another small peak at higher degree. The peak for smaller degrees observed in Fig.~\ref{p_dist_omega} is surprisingly missing.

\begin{figure}
\begin{center}
\includegraphics[width=.9\columnwidth]{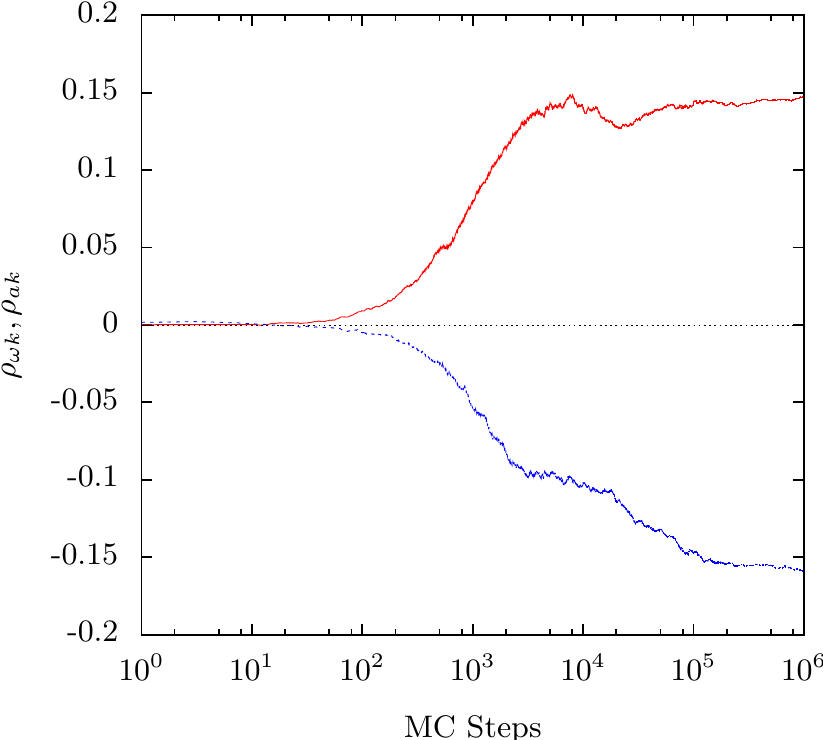}
\end{center}
\caption{\label{rho_omegak_ak}(a) The correlation coefficients $\rho_{\omega k}$ (solid red line) and $\rho_{a k}$ (dotted blue line) are potted as a function of Monte Carlo steps for $64$ coupled R\"ossler systems. The coupled R\"ossler systems have two NDPs, $\omega$ and $a$. Other parameter values are $b=0.2,\ c=7.0$.}
\end{figure}

Now to investigate which nodes are selected as hubs in the optimized network we determine the correlation coefficient $\rho_{\omega k}$ between the node parameter $\omega$ and the node degree $k$ and the correlation coefficient $\rho_{a k}$ between the node parameter $a$ and node degree $k$ using Eq.~(\ref{param-deg-corr}). Fig.~\ref{rho_omegak_ak} shows the correlation coefficients $\rho_{\omega k}$ (solid red line) and $\rho_{a k}$ (dotted blue line) as a function of Monte Carlo steps. The correlation coefficient $\rho_{\omega k}$ increases from zero and saturates to a positive value while the correlation coefficient $\rho_{a k}$ decreases from zero and saturates to negative value. This implies that the nodes with higher value in parameter $\omega$ and lower value in parameter $a$ have larger degree and selected as hubs of the optimized networks.

\begin{figure}
\begin{center}
\includegraphics[width=.9\columnwidth]{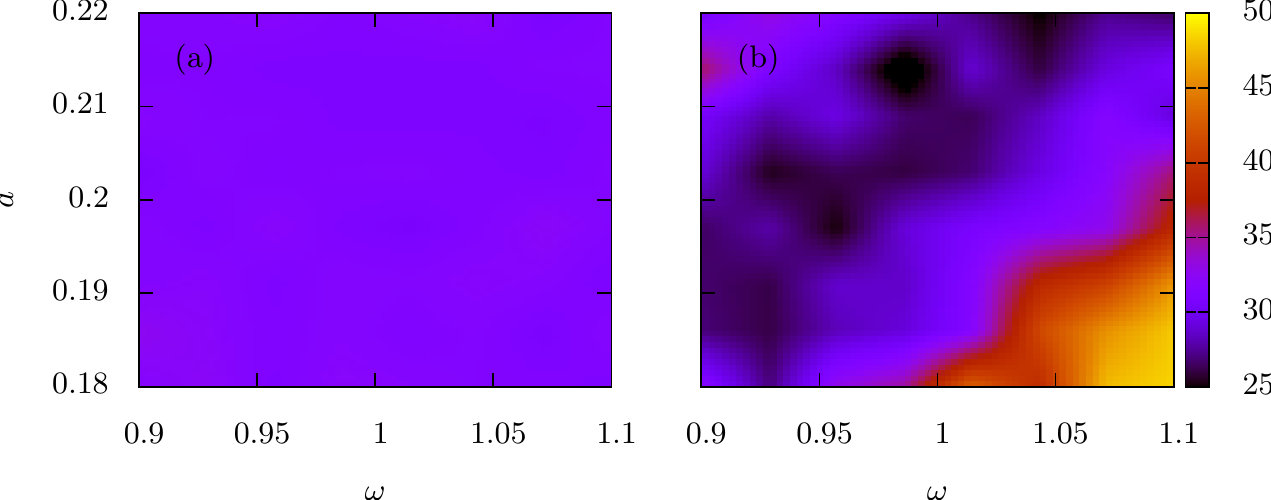}
\end{center}
\caption{\label{Degree_omega_a} The degrees of the nodes of the optimized networks are plotted as grayscale in the parameter plane $(\omega,a)$ for $64$ coupled nonidentical R\"ossler systems.}
\end{figure}

The above conclusions are further supported by Fig.~\ref{Degree_omega_a}, where the degree of the nodes of the optimized networks are plotted as grayscale in the parameter plane $(\omega,a)$ for both random and optimized networks. From the figure we can see that the nodes with higher value in parameter $\omega$ and lower value in parameter $a$ have higher degree than other nodes (bottom right part of the figure). We can also find that there are few nodes (top left part of the figure) with lower value in $\omega$ and higher value in $a$, have relatively higher degree.

\begin{figure}
\begin{center}
\includegraphics[width=.9\columnwidth]{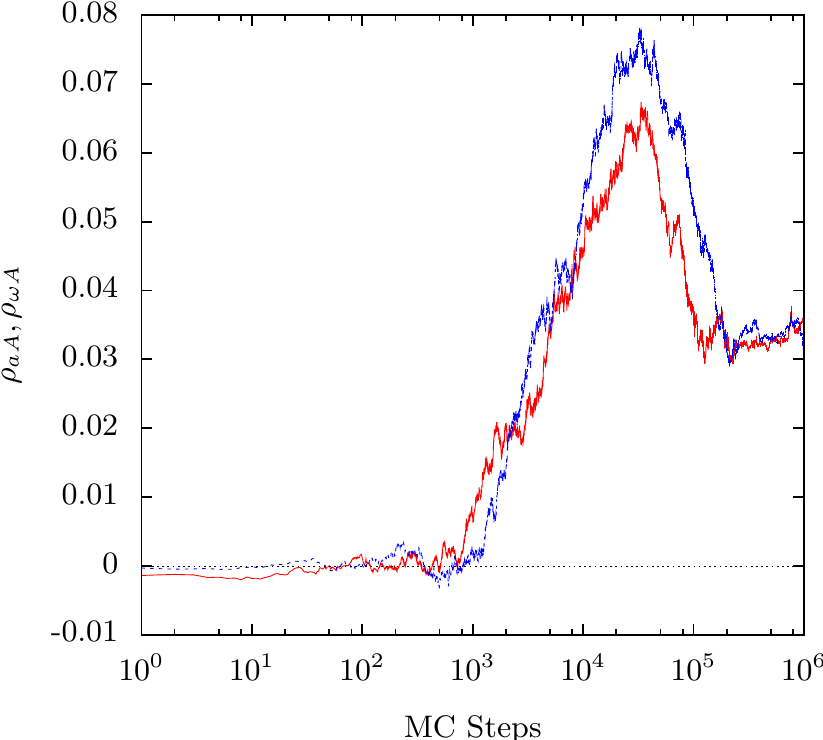}
\end{center}
\caption{\label{rho_omegaA_aA}(a) The correlation coefficients $\rho_{\omega k}$ (solid red line) and $\rho_{a k}$ (dotted blue line) are potted as a function of Monte Carlo steps for $64$ coupled R\"ossler systems. The coupled R\"ossler systems have two NDPs, $\omega$ and $a$. Other parameter values are $b=0.2,\,c=7.0$.}
\end{figure}

\subsection{Links between nodes}

Now we investigate which links are preferable in the optimized networks. We determine two correlation coefficients $\rho_{\omega A}$ and $\rho_{a A}$ (Eq.~(\ref{param-deg-corr})) which give the correlation between the absolute NDP difference for a pair of nodes and the corresponding element of the adjacency matrix $A_{ij}$. In Fig.~\ref{rho_omegaA_aA} the correlation coefficients  $\rho_{\omega A}$ and $\rho_{a A}$ are plotted as a function of Monte Carlo steps. Both the correlation coefficients increases from zero and saturates to positive values. Thus the pairs of nodes which have larger parameter differences in NDPs $\omega$ and $a$ are preferred to create links for the optimized networks.

\subsection{Betweenness centrality and closeness centrality}

\begin{figure}
\begin{center}
\includegraphics[width=.9\columnwidth]{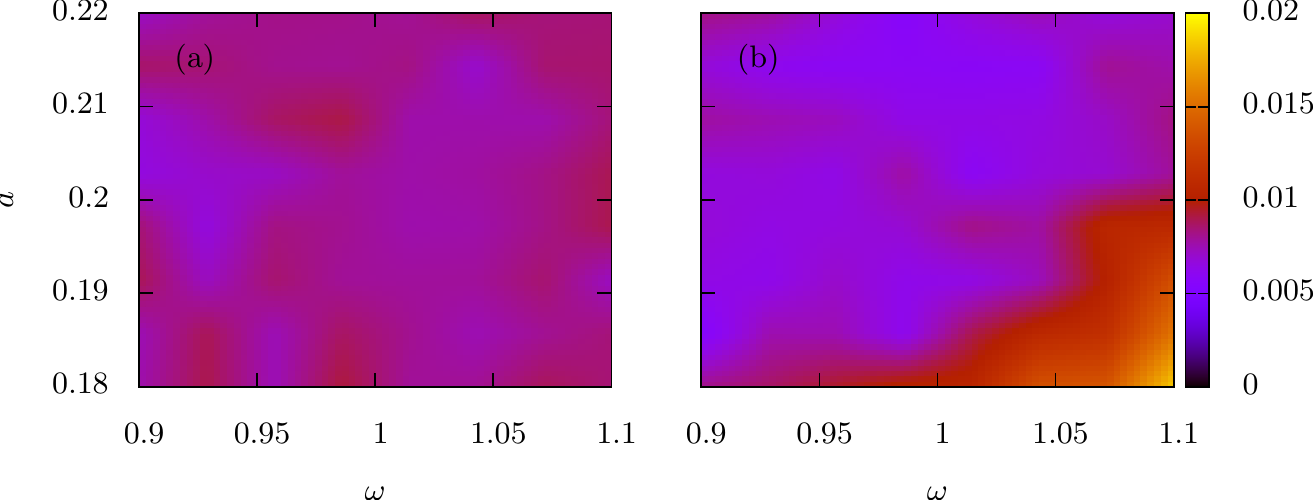}
\end{center}
\caption{\label{BetweennessCentrality_omega_a} The betweenness centrality $C_B$ of the nodes are plotted in grayscale in the parametric plane $(\omega,a)$.}
\end{figure}

Now to investigate which nodes have important role in intra network information transfer, we calculate the betweenness centrality $C_B$ of the nodes. In Figs.~\ref{BetweennessCentrality_omega_a}(a) and (b)  the betweenness centrality of the nodes of the random and optimized networks are plotted as grayscale in the parameter plane $(\omega,a)$. The nodes with higher $\omega$ value and lower $a$ value have higher betweenness centrality $C_B$ than the other nodes of the network.

\begin{figure}
\begin{center}
\includegraphics[width=.9\columnwidth]{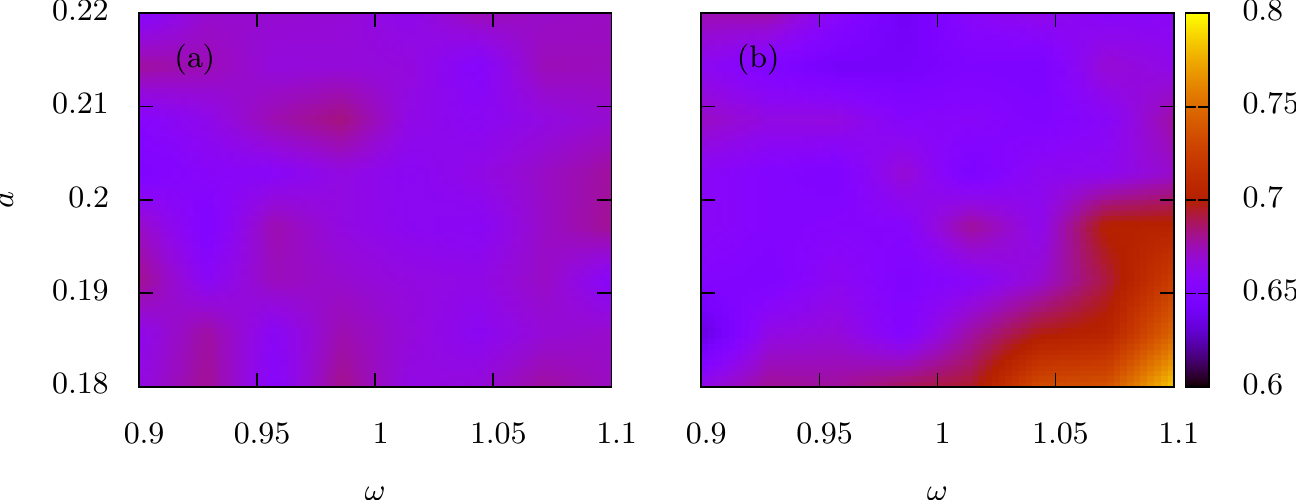}
\end{center}
\caption{\label{ClosenessCentrality_omega_a} The closeness centrality $C_C$ of the nodes of the optimized networks are as grayscale in the parametric plane $(\omega,a)$.}
\end{figure}

In Fig.~\ref{ClosenessCentrality_omega_a}(a) and (b) the closeness centrality $C_C$ of the nodes of the random and optimized networks are plotted as grayscale in the parameter plane $(\omega,a)$. The nodes with higher $\omega$ and lower $a$ have higher closeness centrality $C_C$.

The clustering coefficient, the shortest path lengths and the degree mixing coefficient for two NDPs show a similar behavior to that of one NDP as discussed in the previous Sec. III and hence their plots are not given.

\section{\label{directed-network} Directed networks}

So far we have considered undirected networks. Thus, the coupling matrix is symmetric and all the eigenvalues of the coupling matrix are real and hence one needs to consider only real values of the parameters $\alpha$ and $\nu$. In this section we consider the case where the coupled systems are on directed networks. For a directed network the eigenvalues and the eigenvectors of the coupling matrix can be complex. So, both $\alpha$ and $\nu$ of Eq.~(\ref{master-stability-equation}) can be complex.

\begin{figure}
\begin{center}
\includegraphics[width=.9\columnwidth]{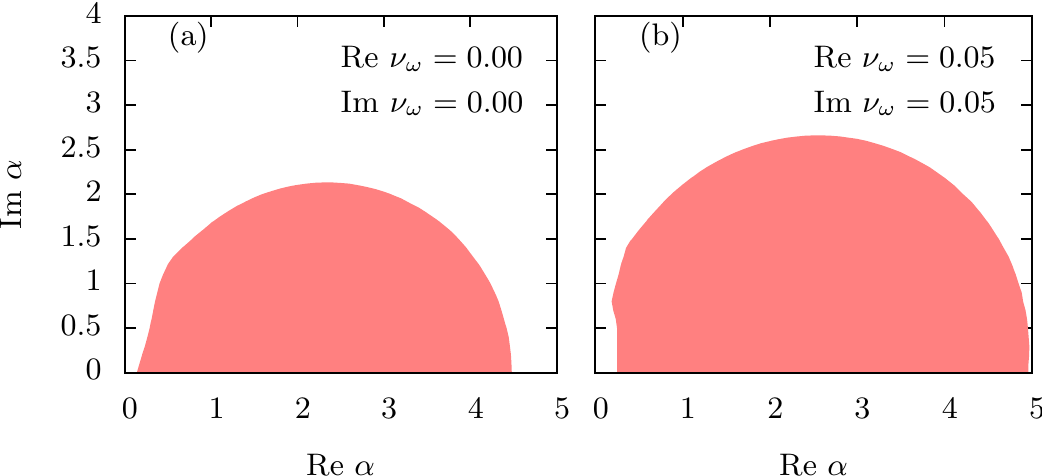}
\end{center}
\caption{\label{a_b_DRe_DIm} The stable regions of synchronization (shaded) are shown in complex plane of parameter $\alpha$ for some chosen discrete values of the complex mismatch parameter $\nu$. In the shaded regions the MSF is negative.}
\end{figure}

We consider the coupled R\"ossler oscillators with NDP as $\omega$. The MSF is calculated from Eq.~(\ref{master-stability-equation}) as a function of the complex parameters $\alpha$ and $\nu$. In Fig.~\ref{a_b_DRe_DIm} the stable regions of synchronization (shaded regions) which are characterized by negative values of MSF, are shown in the complex plane of $\alpha$ for some chosen discrete values of the complex mismatch parameter $\nu$.

\begin{figure}
\begin{center}
\includegraphics[width=.9\columnwidth]{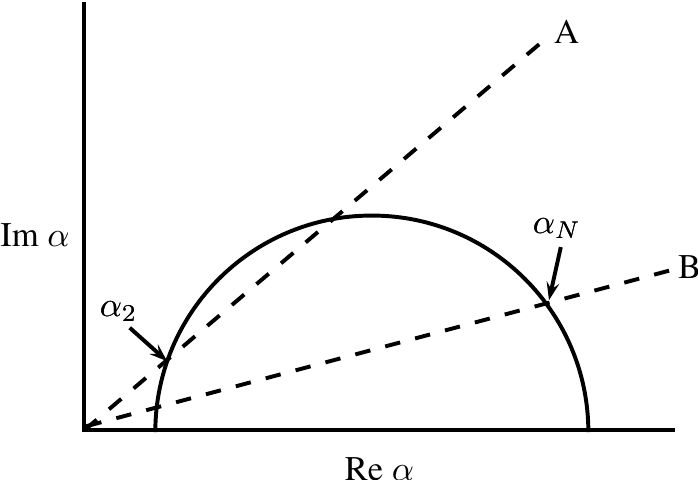}
\end{center}
\caption{\label{msf_zero_contour} The figure shows a schematic plot of the zero contour curves of the MSF for coupled R\"ossler systems in the complex plane of the parameter $\alpha$ for a fixed $\nu$. The lines A and B and the intersection points $\alpha_2$ and $\alpha_N$ are explained in the text.}
\end{figure}

For constructing the optimized networks we follow the before mentioned Metropolis algorithm. For this algorithm, we need the interval $l_{\varepsilon}$ of the coupling parameter $\varepsilon$ which gives stable synchronization. To determine the interval $l_{\varepsilon}$, we use the following procedure. Let the coupling matrix of a directed network be $G$ and let its eigenvalues be $\gamma_i;\;i=1,...,N$, ordered according to the real part of $\gamma_i$. The eigenvalue $\gamma_1=0$ and is related to the synchronization manifold and the other nonzero eigenvalues gives the transverse manifold. The two extreme transverse eigenvalues of $G$ are $\gamma_2$ and $\gamma_N$. The synchronization is stable when all Lyapunov exponents corresponding to the nonzero eigenvalues lie in the negative region of the MSF. In Fig.~\ref{msf_zero_contour}, we show a schematic plot of the zero contour line of MSF for $x$ component coupled R\"ossler oscillators in the complex $\alpha$ plane for a given complex mismatch parameter $\
nu$. The MSF is negative in the region below the zero contour line. The two dashed lines A and B in Fig.~\ref{msf_zero_contour} have slopes equal to $|\rm{Im}\gamma_2|/|\rm{Re}\gamma_2|$
and $|\rm{Im}\gamma_N|/|\rm{Re}\gamma_N|$ respectively. Both the lines A and B cut the zero contour of the MSF at two points. For line A we choose the first cut and call it $\alpha_2$ and for line B we choose the second cut and call it $\alpha_N$ \cite{note-bifurcation}. The stable interval $l_{\varepsilon}$ is determined as, $l_{\varepsilon} = |\rm{Re}\alpha_2/\rm{Re}\gamma_2 - \rm{Re}\alpha_N/\rm{Re}\gamma_N|$. This procedure to determine $l_{\varepsilon}$ is for a given value of $\nu$. To determine $l_{\varepsilon}$ for an arbitrary $\nu$ using the fixed chosen values of $\nu$ as in Fig.~\ref{a_b_DRe_DIm}, we use linear interpolation.

\begin{figure}
\begin{center}
\includegraphics[width=.9\columnwidth]{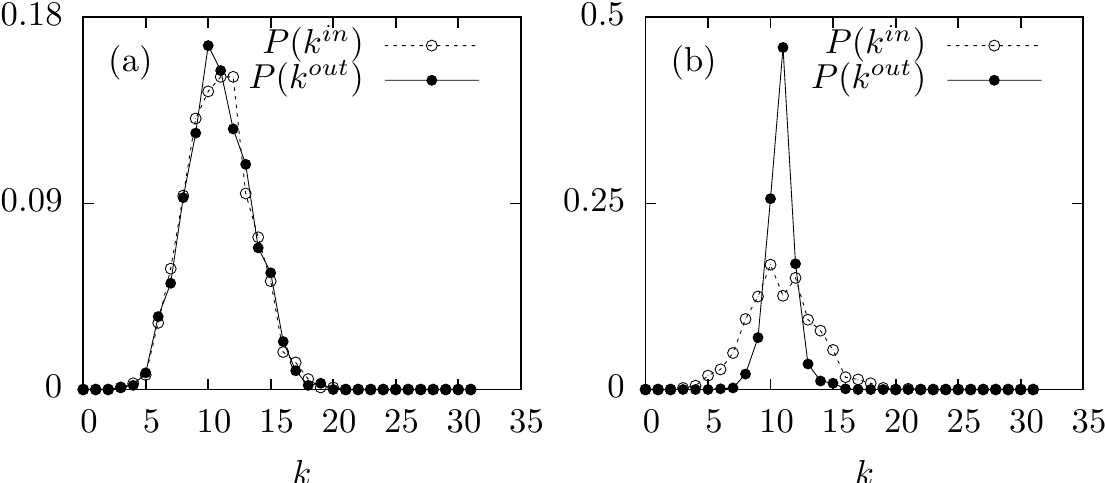}
\end{center}
\caption{\label{OutDegreeDist} (a) The in-degree distribution $P(k^{in})$ is shown as a function of $k^{in}$ for the random (dashed line) and optimized networks (solid line). (b) The out-degree distribution $P(k^{out})$ is shown as a function of $k^{out}$ for the random (dashed line) and optimized networks (solid line).
}
\end{figure}

For directed networks each node has an in-degree and an out-degree and hence a separate degree distribution for each is required.
In Fig.~\ref{InDegreeDist}(a) we plot the in-degree distribution $P(k^{in})$ as a function of $k^{in}$ for the random (dashed line) and optimized networks (solid line) and in Fig.~\ref{OutDegreeDist}(b) we show a similar plot for the out-degree distribution $P(k^{out})$. For both in- and out-degrees the initial distributions for random network are gaussian.  For the optimized network the in-degree distribution is almost similar to random case while the out-degree distribution is considerably broadened.


\begin{figure}
\begin{center}
\includegraphics[width=.9\columnwidth]{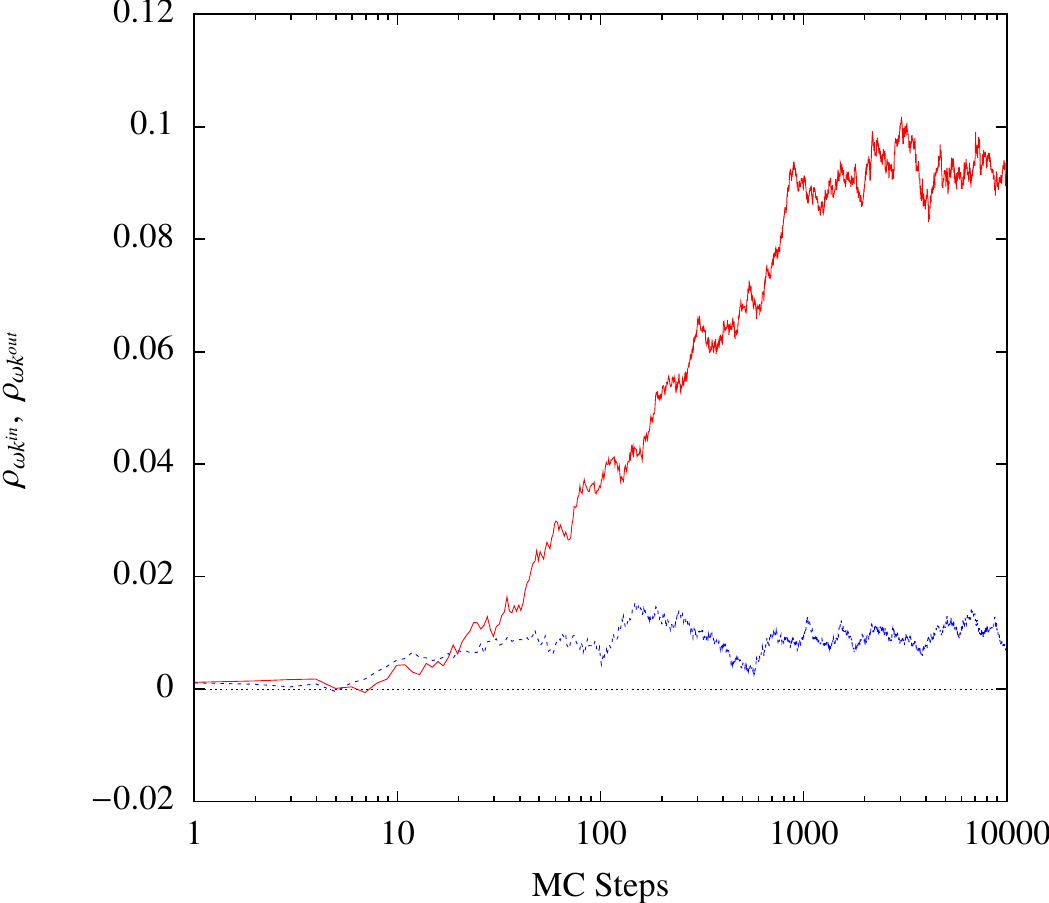}
\end{center}
\caption{\label{rho_omegakin_omegakout} The correlation coefficients $\rho_{\omega k^{in}}$ (red curve) and $\rho_{\omega k^{out}}$ (blue curve) are shown as a function of the Monte Carlo steps. The correlation coefficient $\rho_{\omega k^{in}}$ increases from zero and saturates to a positive value, but $\rho_{\omega k^{out}}$ remains close to zero.}
\end{figure}

To understand the degree distributions, we look for separate correlations between the in- and out-degrees and the corresponding NDP, $\rho_{\omega k^{in}}$ and $\rho_{\omega k^{out}}$ which are defined as in Eq.~(\ref{param-deg-corr}).
In Fig.~\ref{rho_omegakin_omegakout}, the correlation coefficients $\rho_{\omega k^{in}}$ (red curve) and $\rho_{\omega k^{out}}$ (blue curve) are shown as a function of the Monte Carlo steps. $\rho_{\omega k^{in}}$ increases from zero and saturates to positive value, while $\rho_{\omega k^{out}}$ remains near zero. So, in the optimized networks the nodes with higher $\omega$ value have larger in-degree $k^{in}$, while the out-degrees $k^{out}$ does not show any correlation with $\omega$.

\begin{figure}
\begin{center}
\includegraphics[width=.9\columnwidth]{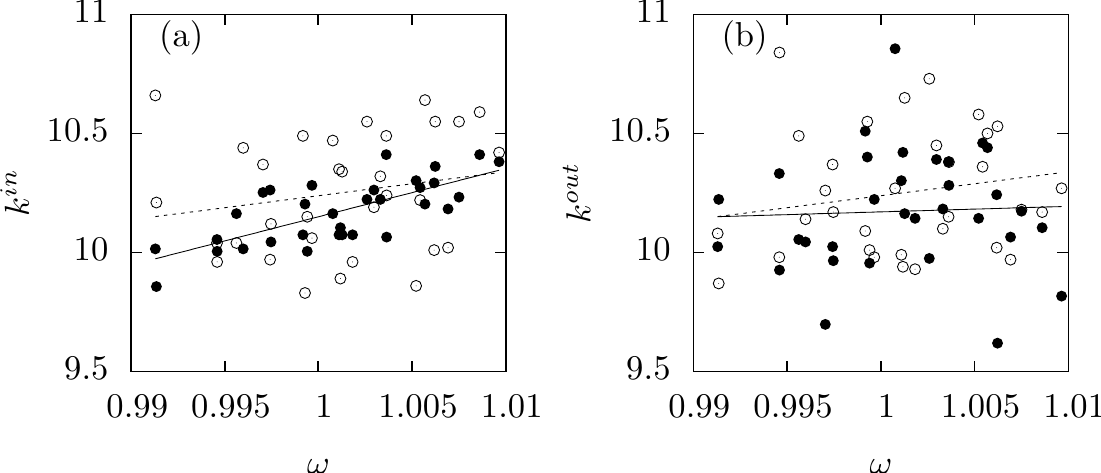}
\end{center}
\caption{\label{omega_kin_omega_kout}(a) The in-degree $k^{in}$ of the nodes of the random (open circles) and optimized (solid circles) networks are plotted as a function of NDP $\omega$. (b) the out-degree $k^{out}$ of the nodes of the random (open circles) and optimized (solid circles) networks are plotted as a function of the NDP $\omega$.}
\end{figure}

We plot the in-degree $k^{in}$ for the random (open circles) and optimized (solid circles) networks as a function of the NDP $\omega$ in Fig.~\ref{omega_kin_omega_kout}(a) and have a similar plot for the out-degree $k^{out}$ in Fig.~\ref{omega_kin_omega_kout}(b). From Fig.~\ref{omega_kin_omega_kout}(a) we can see that the node with higher $\omega$ have higher in-degree, while Fig.~\ref{omega_kin_omega_kout}(b) shows that there in so such correlation between the out-degree and $\omega$. This result is in agreement with the correlation coefficient plots in Fig.~\ref{rho_omegakin_omegakout}.

\section{\label{conclusion} Conclusion}
Using the MSF for nearly identical systems as derived in \cite{previous}, we construct a synchronization optimized network by rearranging the edges of a given network. We then study the relations between the different network properties and the corresponding distribution of NDP, for the optimized network and compare these relations with those of the random network. In the optimized network, we find that the nodes with parameter value at one end of the NDP distribution are chosen as hubs, and also they have higher betweenness centrality. Thus, these nodes play a key role in the information transfer between the coupled systems. We have found that the pair of nodes with larger parameter difference are preferred to create links in the optimized networks. The clustering coefficient of the optimized networks are higher than the clustering coefficient of the starting random networks. We have also studied other properties such as clustering coefficient, average shortest path length, closeness centrality and degree 
mixing and find similar trends in their relation with the NDP.

\section{\label{acknowledgements}Acknowledgements}

All the numerical calculations are done on the high performance computing clusters at PRL.

\end{document}